\documentclass[useAMS,usenatbib]{mn2e}

\usepackage{graphicx}
\usepackage{epsfig,floatflt}

\title[Spectroscopic binarity of X~Pup and XX~Sgr]
{Discovery of the spectroscopic binary nature 
of the Cepheids X~Puppis and XX~Sagittarii}
\author[L. Szabados et~al.]{L\'aszl\'o Szabados$^{1}$\thanks{E-mail:
szabados@konkoly.hu}, Aliz Derekas$^{1}$, Csaba Kiss$^{1}$, and
P\'eter Klagyivik$^{1}$
\\
$^{1}$Konkoly Observatory,
Research Centre for Astronomy and Earth Sciences,
Hungarian Academy of Sciences,\\
H-1121 Budapest, Konkoly Thege Mikl\'os \'ut 15-17, Hungary\\}
\begin{document}

\date{Accepted Received ; in original form }

\pagerange{\pageref{firstpage}--\pageref{lastpage}} \pubyear{2012}

\maketitle

\label{firstpage}

\begin{abstract}
We present the analysis of photometric and spectroscopic data 
of two bright Galactic Cepheids, X Puppis and XX Sagittarii. 
Based on the available data in the literature as well as own 
observations spanning 75 years, we conclude that both Cepheids 
belong to spectroscopic binary systems.
However, the data are not sufficient to determine the orbital 
periods nor other elements for the orbit. This discovery 
corroborates the statement on the high frequency of occurrence 
of binaries among the classical Cepheids, a fact to be taken into 
account when calibrating the period-luminosity relationship for 
Cepheids. The photometric data revealed that the pulsation period 
of X~Pup is continuously increasing with $\Delta P={\rm 0.007559
d/century}$ likely caused by stellar evolution. The pulsation period 
of XX~Sgr turned out to be very stable in the last $\sim$100 years.

\end{abstract}

\begin{keywords}
stars: variables: Cepheids -- binaries: spectroscopic
\end{keywords}

\section{Introduction}
\label{intro}

Classical Cepheid variable stars are key objects in astronomy
because owing to their radial pulsation and its consequences--mainly 
the famous period-luminosity ($P$-$L$) relationship--they 
rank among standard candles in establishing the cosmic distance 
scale and serve as test objects of stellar evolution of
intermediate-mass stars. 

Companions to Cepheids, however, complicate the situation.
On the one hand, the contribution of the secondary star to the
observed brightness has to be taken into account when involving any
particular Cepheid in the calibration of the $P$-$L$ relationship
and the evolution of binary stars may be quite different from
the single star evolution (depending on the separation of the
components). On the other hand, frequency of binaries (and
multiple stars) among classical Cepheid variables is considerable:
it exceeds 50 per cent for brightest Cepheids \citep{Sz03a}, 
while among the fainter Cepheids an observational selection 
effect encumbers revealing binarity. In a broader aspect, however,
this frequent occurrence of binaries among Cepheids is not
surprising, it agrees with the general frequency of binary stars
in the solar neighbourhood: two-third of solar type stars of 
F3-G2 spectral type stars have stellar companions \citep{AL76}.

Depending on the brightness and temperature differences between
the Cepheid and its (either optical or physical) companion, the
observable brightness and colour of the unresolved binary
system can differ from the respective value intrinsic to the
Cepheid. If uncorrected for its contribution, the companion
can falsify the luminosity and radius of the Cepheid derived by
using the Baade-Wesselink method (except its infrared surface 
brightness method implementation). 

The orbital motion of a Cepheid 
in a binary system can even lead to a wrong trigonometric parallax
if no allowance is made for binarity. An illustrative example
for this adverse effect was shown by \citet{Szetal11}: all 
negative Hipparcos values for Cepheids within 2 kpc were solely 
derived in the case of Cepheids with known spectroscopic companions.

Cepheids belonging to open clusters are often used for 
calibrating the $P$-$L$ relationship based on the independently 
derived cluster distances \citep[e.g.,][]{TB02,T10}. 
For these calibrating Cepheids it is especially important to correct 
for the luminosity contribution from the companion.

Therefore, it is essential to study Cepheids individually 
from the point of view of binarity before involving them in any
calibration procedure (of e.g., $P$-$L$ or period-radius
relationship).

Hot companions to Cepheids can be effectively discovered by
ultraviolet {\em spectroscopy}: the {\it IUE\/} satellite was instrumental
in revealing early type companions \citep{E92}. Radial velocity
time series data (normally in the optical region) obtained during 
at least two widely differing epochs can lead to discovery of
spectroscopic binaries \citep[see e.g.,][]{Sz96}.
Revealing binarity by means of {\em astrometry} will be available from
the data to be obtained during the ESA {\it Gaia\/} space mission
from 2013 on.
Owing to the regular behaviour of Cepheid pulsation, various 
{\em photometric} criteria have also been devised for pointing out
the presence of Cepheid companions \citep{Sz03b,KSz09}.

In this paper we reveal {\em spectroscopic binarity} of two bright
Cepheids, X~Puppis and XX~Sagittarii, by carefully analysing 
the radial velocity data published on these variable stars.
Some new radial velocity data obtained for XX~Sgr have also been 
included in the analysis.

In the case of pulsating variables, like Cepheids, spectroscopic
binarity manifests itself in a periodic variation of the
$\gamma$-velocity (i.e., the radial velocity of the mass centre
of the Cepheid). In practice, the orbital radial velocity variation
of the Cepheid component is superimposed on the radial velocity
variations of pulsational origin. To separate the orbital and 
pulsational effects, knowledge of the accurate pulsation period
is essential, especially when comparing radial velocity
data obtained at widely differing epochs. Therefore, the pulsation
period and its variations have been determined with the method of
the $O-C$ diagram \citep{S05} for both target Cepheids. Use of the
accurate pulsation period obtained from the photometric data is a 
guarantee for the correct phase matching of the (usually less
precise) radial velocity data.

%%%%%%%%%%%%%%%
\section{X~Puppis}
\label{xpup}

\subsection{Accurate value of the pulsation period}
\label{xpup-period}

The brightness variability of X~Pup (HD\,60266, $\langle V \rangle
= 8.46$\,mag.) was revealed by \citet{K90}.
During the time interval elapsed since the discovery (spanning 
more than 120 years) the photometric variability was followed 
first visually, then photographically, from the 1950s
photoelectrically, and in the last decades by CCD photometry. 
All published observations of this Cepheid radially pulsating 
in the fundamental mode were re-analysed in a homogeneous 
manner to determine seasonal moments of the normal maxima.
These data are collected in Table~\ref{tab-xpup-oc} whose
column contain the following pieces of information:\\
Column~1: Heliocentric normal maxima;\\
Col.~2: Epoch number, $E$, as calculated from Eq.~1:
\vspace{-1mm}
\begin{equation}
C = 2\,454\,845.0497 + 25.970\,068{\times}E 
\label{xpup-ephemeris}
\end{equation}
\vspace{-3mm}
$\phantom{mmmmm}\pm0.0926\phantom{\,}\pm 0.000\,406$

\noindent (this ephemeris has been obtained by the weighted least squares
fit to the tabulated $O-C$ residuals);\\
\noindent Col.~3: the corresponding $O-C$ residual as calculated 
from Eq.~\ref{xpup-ephemeris};\\
Col.~4: weight assigned to the $O-C$ residual (1, 2, or 3 
depending on the quality of the light curve leading to 
the given residual);\\
Col.~5: reference to the origin of data, preceded by the name 
of the observer if different from the author(s) cited.\\

\begin{figure}
\includegraphics[height=70mm, angle=0]{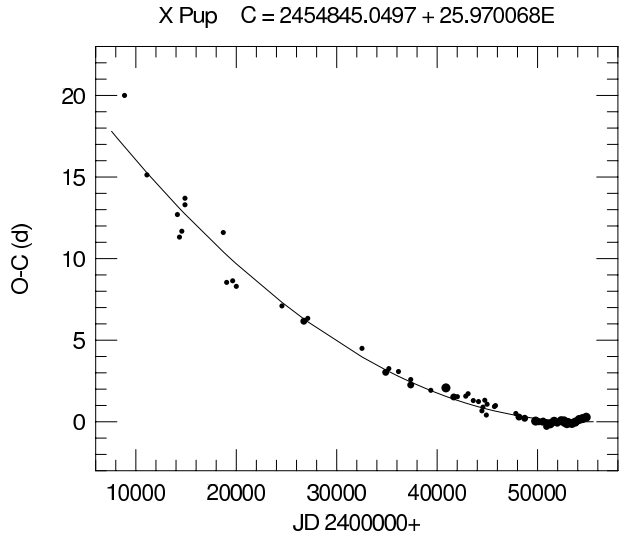}
\caption{$O-C$ diagram of X~Pup based on the residuals
listed in Table~\ref{tab-xpup-oc}. The pulsation period
of X~Pup is continuously increasing}
\label{fig-xpup-oc}
\end{figure}

\begin{figure}
\includegraphics[height=57mm, angle=0]{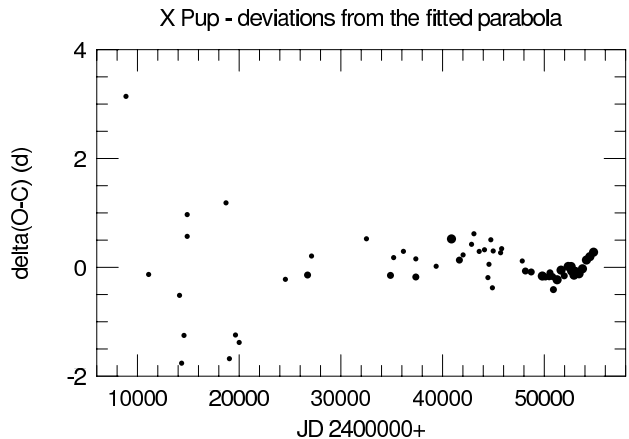}
\caption{Deviations of the $O-C$ residuals of X~Pup from 
the fitted parabola}
\label{fig-xpup-deltaoc}
\end{figure}

The $O-C$ residuals have been plotted in Fig.~\ref{fig-xpup-oc}
together with the least squares fitted parabola. This
parabolic trend corresponds to a continuous period increase
of $5.375{\times}10^{-6}$ d/cycle, i.e., $\Delta P = 0.007559$
d/century. This tiny but non-negligible effect has been caused
by stellar evolution: while the Cepheid crosses the instability
region towards lower temperatures in the Hertzsprung-Russell 
diagram, its pulsation period is increasing. Superimposed on
this parabolic trend, random fluctuations of the period value
are also seen on a time scale of several years -- see the
quasi continuously covered part of the deviations from the
parabolic fit to the $O-C$ residuals after JD\,2\,448\,000 
in Fig.~\ref{fig-xpup-deltaoc}. The amplitude of these
fluctuations is several tenths of a day, an order of magnitude 
larger than the periodic light-time effect expected in
a binary system (of suitably placed orbital plane) with a 
Cepheid component.

%%%%Fig. 1

%%%Fig. 2

%%%%%%%%%%%%Table 1
\begin{table}
\caption{$O-C$ residuals of X~Puppis (description of the columns
is given in Sect.~\ref{xpup-period})}
\begin{tabular}{l@{\hskip2mm}r@{\hskip2mm}r@{\hskip2mm}c@{\hskip2mm}l}
\hline
\noalign{\vskip 0.3mm}
JD$_{\odot}$ & $E\ $ & $O-C$ & $W$ & Data source\\
2\,400\,000 + &&&\\
\noalign{\vskip 0.3mm}
\hline
\noalign{\vskip 0.3mm}
08872      & $-$1771 & 20     & 1 & Sch\"onfeld \citep{P56}\\
11100.62   & $-$1685 & 15.13  & 1 & \citet{K90}\\
14136.7    & $-$1568 & 12.7   & 1 & Perry \citep{P56}\\
14343.0631 & $-$1560 & 11.3195& 1 & \citet{Z32}\\
14577.1534 & $-$1551 & 11.6792& 1 & \citet{P97,P99,P03}\\
14890.4    & $-$1539 & 13.3   & 1 & Hartwig \citep{P56}\\
14890.8    & $-$1539 & 13.7   & 1 & Innes \citep{P56}\\
18706.3    & $-$1392 & 11.6   & 1 & Worsell \citep{P56}\\
19040.869  & $-$1379 & 8.543  & 1 & Robinson \citep{P56}\\
19638.28   & $-$1356 & 8.64   & 1 & \citet{H28}\\
20001.52   & $-$1342 & 8.30   & 1 & Payne \& Gaposchkin\\
&&&& \citep{P56}\\
24545.1    & $-$1167 & 7.1    & 1 & \citet{B33}\\
26725.6293 & $-$1083 & 6.1632 & 2 & \citet{O'C34}\\
27115.3559 & $-$1068 & 6.3388 & 1 & \citet{FK53}\\
32515.29    & $-$860 & 4.50   & 1 & Berdnikov, Mattei \& \\
            &        &        &   & Beck (2003)\\
34877.0978  & $-$769 & 3.0304 & 2 & Walraven, Muller \&\\
            &        &        &   & Oosterhoff (1958)\\
35188.9673  & $-$757 & 3.2591 & 1 & \citet{I61}\\
36149.68    & $-$720 & 3.08   & 1 & \citet{Betal03}\\
37369.7691  & $-$673 & 2.2572 & 2 & \citet{Metal64}\\
37369.78    & $-$673 & 2.59   & 1 & \citet{Betal03}\\
39368.8181  & $-$596 & 1.9289 & 1 & \citet{T69}\\
40875.2311  & $-$538 & 2.0780 & 3 & \citet{P76}\\
41653.7755  & $-$508 & 1.5203 & 2 & \citet{M75}\\
42017.38    & $-$494 & 1.54   & 1 & \citet{Betal03}\\
42848.4488  & $-$462 & 1.5705 & 1 & \citet{D77}\\
43082.33    & $-$453 & 1.72   & 1 & \citet{Betal03}\\
43601.31    & $-$433 & 1.30   & 1 & \citet{Betal03}\\
44120.65    & $-$413 & 1.24   & 1 & \citet{Betal03}\\
44457.6939  & $-$400 & 0.6714 & 1 & \citet{E83}\\
44561.8016  & $-$396 & 0.8988 & 1 & \citet{B08}\\
44744.01    & $-$389 & 1.32   & 1 & \citet{Betal03}\\
44898.9264  & $-$383 & 0.4127 & 1 & \citet{MB84}\\
44977.5009  & $-$380 & 1.0770 & 1 & \citet{B08}\\
45704.52    & $-$352 & 0.9342 & 1 & \citet{Betal03}\\
45808.46    & $-$348 & 0.9940 & 1 & \citet{Betal03}\\
47833.64    & $-$270 & 0.5087 & 1 & \citet{Betal03}\\
48145.0626  & $-$258 & 0.2904 & 2 & Hipparcos \citep{ESA97}\\
48716.3278  & $-$236 & 0.2141 & 2 & Hipparcos \citep{ESA97}\\
49806.8990  & $-$194 & 0.0425 & 3 & \citet{B08}\\
50118.4982  & $-$182 & 0.0009 & 2 & \citet{B02}\\
50482.0566  & $-$168 & $-$0.0217 & 2 & \citet{B02}\\
50560.0343  & $-$165 & 0.0457 & 2 & \citet{B08}\\
50819.6483  & $-$155 & $-$0.0409 & 2 & \citet{B08}\\
50897.3144  & $-$152 & $-$0.2850 & 2 & \citet{B08}\\
51261.0546  & $-$138 & $-$0.1257 & 3 & \citet{B08}\\
51650.7613  & $-$123 & 0.0300 & 3 & \citet{B08}\\
51962.2826  & $-$111 & $-$0.0896 & 2 & \citet{B08}\\
52351.9893   & $-$96 & 0.0661 & 3 & \citet{B08}\\
52637.6479   & $-$85 & 0.0540 & 3 & \citet{B08}\\
52689.5032   & $-$83 & $-$0.0309 & 3 & ASAS \citep{P02}\\
52923.1507   & $-$74 & $-$0.1140 & 3 & ASAS \citep{P02}\\
53001.1284   & $-$71 & $-$0.0465 & 3 & ASAS \citep{P02}\\
53001.1284   & $-$71 & $-$0.0465 & 3 & \citet{B08}\\
53416.5940   & $-$55 & $-$0.1020 & 3 & ASAS \citep{P02}\\
53754.2896   & $-$42 & $-$0.0172 & 3 & ASAS \citep{P02}\\
54143.9963   & $-$27 & 0.1384 & 3 & ASAS \citep{P02}\\
54481.6660   & $-$14 & 0.1973 & 3 & ASAS \citep{P02}\\
54845.3282   &  0    & 0.2785 & 3 & ASAS \citep{P02}\\
\noalign{\vskip 0.3mm}
\hline
\end{tabular}
\label{tab-xpup-oc}
\end{table}

%%%%%%%%%%%%%%%%
\subsection{Radial velocity data of X~Pup}
\label{xpup-radvel}

%%%%%%%Table 2
\begin{table}
\caption{$\gamma$-velocities of X~Puppis}
\begin{tabular}{lccl}
\hline
\noalign{\vskip 0.3mm}
Mid-JD & $v_{\gamma}$ & $\sigma$ & Data source \\
2\,400\,000+ & (km/s)& (km/s) & \\
\noalign{\vskip 0.3mm}
\hline
\noalign{\vskip 0.3mm}
25590 & 65.0 & 2.0 & \cite{J37}\\
44666 & 67.4 & 0.5 & \cite{Cetal01} \\
45065 & 66.5 & 1.5 & \cite{Betal88} \\
50523 & 70.0 & 0.4 & \cite{B02} \\
51111 & 71.5 & 0.3 & \cite{Petal05} \\
54922 & 71.8 & 0.2 & \cite{Setal11}\\
\noalign{\vskip 0.3mm}
\hline
\end{tabular}
\label{tab-xpup-vgamma}
\end{table}

%%%Fig. 3
\begin{figure}
\includegraphics[height=55mm, angle=0]{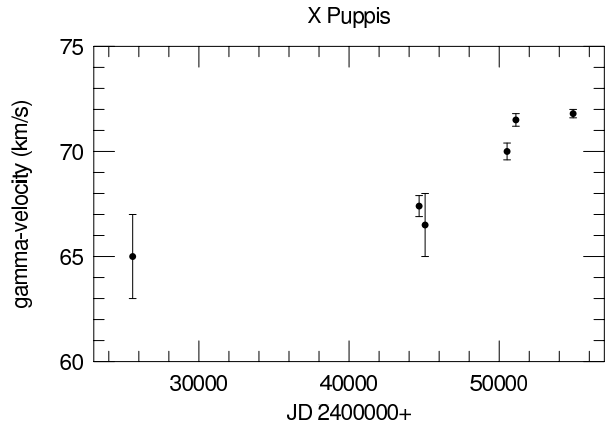}
\caption{Temporal drift in the $\gamma$-velocity
of X~Puppis}
\label{fig-xpup-vrad}
\end{figure}

The available radial velocity data include those obtained by
\cite{J37}, \cite{Cetal01}, Barnes, Moffett \& Slovak (1988), 
\cite{B02}, \cite{Petal05}, and \cite{Setal11}. The phase curve was
constructed from each data set using the actual value of the
pulsation period taking into account the continuous period
variation implied by Fig.~\ref{fig-xpup-oc}.
 (see Sect.~\ref{xpup-period}). Then the mean
value of the radial velocity (the $\gamma$-velocity)
was determined for each data set. These $\gamma$-velocities
(together with their uncertainties are listed in 
Table~\ref{tab-xpup-vgamma} and also plotted in 
Fig.~\ref{fig-xpup-vrad}. The pattern of the data points
implies a monotonically changing $\gamma$-velocity which is a 
sign of the orbital motion in a spectroscopic binary system.
The orbital period can be as long as several decades.

Spectroscopic binarity of X~Pup has to be confirmed by further
observations because the earliest data (obtained by Joy about
80 years ago) are of low quality. Nevertheless, all radial
velocity data have been obtained based on the observed wavelength
of metallic lines, therefore one does not expect a noticeable
systematic difference between the various data sets.

%%%%%%%%%%%%%
\section{XX~Sagittarii}
\label{xxsgr}

\subsection{Accurate value of the pulsation period}
\label{xxsgr-period}

The brightness variability of XX~Sagittarii (HD\,169315,
$\langle V \rangle = 8.65$\,mag.) was discovered by Annie Cannon
\citep{P08}. In the first half of the 20th century XX~Sgr
was occasionally observed mainly visually, regular 
photometric data on this Cepheid are available from the
last three decades. All photometric data of this single
periodic Cepheid pulsating in the radial fundamental mode
have been subjected to an $O-C$ analysis whose results are 
summarized in Table~\ref{tab-xxsgr-oc}. The $O-C$ residuals in 
this table have been obtained by using the following final
ephemeris:

\vspace{-1mm}
\begin{equation}
C = 2\,452\,814.4629 + 6.424\,267{\times}E 
\label{xxsgr-ephemeris}
\end{equation}
\vspace{-3mm}
$\phantom{mmmmm}\pm0.0059\phantom{\,}\pm 0.000\,003$

\noindent determined from a weighted least-squares fit to 
all normal maxima listed in Table~\ref{tab-xxsgr-oc}.

\begin{figure}
\includegraphics[height=45mm, angle=0]{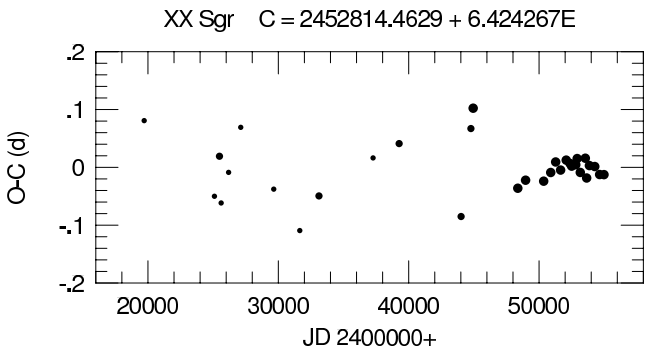}
\caption{$O-C$ diagram of XX~Sgr. The plot can be
approximated by a constant period}
\label{fig-xxsgr-oc}
\end{figure}

The $O-C$ diagram of XX~Sgr is plotted in Fig.~\ref{fig-xxsgr-oc}. 
The pattern of data can be approximated by a constant period
but in the last decades a wave-like pattern is seemingly
superimposed on the $O-C =0$ constant line. However, this 
feature, can hardly be attributed to a light-time effect
occurring in a binary system due to the orbital motion
because the amplitude of the $O-C$ variations is too large.

%%%%%%%%%%%%Table 3
\begin{table}
\caption{$O-C$ residuals of XX~Sagittarii}
\begin{tabular}{l@{\hskip2mm}r@{\hskip2mm}r@{\hskip2mm}c@{\hskip2mm}l}
\hline
\noalign{\vskip 0.3mm}
JD$_{\odot}$ & $E\ $ & $O-C$ & $W$ & Data source\\
2\,400\,000 + &&&\\
\noalign{\vskip 0.3mm}
\hline
\noalign{\vskip 0.3mm}
19710.2958  & $-$5153  & 0.0808 & 1 & Zinner \citep{V32}\\
25100.13    & $-$4314  & $-$0.05   & 1 & \citet{Z29}\\
25485.6503  & $-$4254  & 0.0192 & 2 & \citet{V32}\\
25614.0548  & $-$4234  & $-$0.0616 & 1 & \citet{Z54}\\
26185.8675  & $-$4145  & $-$0.0087 & 1 & \citet{P38}\\
27117.4641  & $-$4000  & 0.0692 & 1 & \citet{FK53}\\
29648.5184  & $-$3606  & $-$0.0377 & 1 & \citet{S48}\\
31646.3938  & $-$3295  & $-$0.1093 & 1 & \citet{S48}\\
33117.6109  & $-$3066  & $-$0.0494 & 2 & \citet{Eetal57}\\
37267.7531  & $-$2420  & 0.0163 & 1 & \citet{Metal64}\\
39265.7249  & $-$2109  & 0.0411 & 2 & \citet{T69}\\
44019.5564  & $-$1369  & $-$0.0850 & 2 & \citet{B08}\\ 
44771.3477  & $-$1252  & 0.0671 & 2 & \citet{MB84}\\
44944.8381  & $-$1225  & 0.1023 & 3 & \citet{B08}\\
48368.8340   & $-$692  & $-$0.0361 & 3 & {\it Hipparcos\/} \citep{ESA97}\\
48966.3048   & $-$599  & $-$0.0222 & 3 & \citet{Aetal98}\\
50353.9447   & $-$383  & $-$0.0239 & 3 & \citet{B08}\\
50893.5982   & $-$299  & $-$0.0089 & 3 & \citet{B08}\\
51272.6481   & $-$240  & 0.0093 & 3 & \citet{B08}\\
51651.6659   & $-$181  & $-$0.0047 & 3 & \citet{B08}\\
52082.1090   & $-$114  & 0.0125 & 3 & ASAS \citep{P02}\\
52371.1953    & $-$69  & 0.0068 & 3 & \citet{B08}\\
52499.6717    & $-$49  & 0.0021 & 3 & ASAS \citep{P02}\\
52814.4674   &   0     & 0.0045 & 3 & ASAS \citep{P02}\\
52923.6906   &  17     & 0.0152 & 3 & {\it INTEGRAL\/} OMC\\
53161.3645   &  54     & $-$0.0088 & 3 & ASAS \citep{P02}\\
53546.8451   & 114     & 0.0158 & 3 & ASAS \citep{P02}\\
53649.5992   & 130     & $-$0.0184 & 3 & {\it INTEGRAL\/} OMC\\
53861.6215   & 163     & 0.0031 & 3 & ASAS \citep{P02}\\
54272.7729   & 227     & 0.0014 & 3 & ASAS \citep{P02}\\
54645.3666   & 285     & $-$0.0124 & 3 & ASAS \citep{P02}\\
54979.4283   & 337     & $-$0.0126 & 3 & ASAS \citep{P02}\\
\noalign{\vskip 0.3mm}
\hline
\end{tabular}
\label{tab-xxsgr-oc}
\end{table}

%%%Fig. 4

%%%%%%%%%%%%%%%%
\subsection{Radial velocity data of XX~Sgr}
\label{xxsgr-radvel}

Over half century the values published by \citet{J37}
had been the only existing radial velocity data on XX~Sgr.
The radial velocity survey of \citet{Betal88} also included
this Cepheid, and their two data implied a $\gamma$-velocity
value different from the one determined Joy's data. 
Suspecting the spectroscopic binary nature we initiated
new radial velocity observations (discussed in Sect.~\ref{feros}).
In the meantime, however, some more radial velocity data
have been published by \citet{Betal10} and \cite{Setal11}.
This latter series of observations resulted in a well-covered
and very accurate radial velocity phase curve which confirms
the spectroscopic binarity of XX~Sgr. However, \cite{Setal11}
did not compare their own measurements with the previous data,
thus they missed to reveal the binarity of this Cepheid.

%%%%%%%%%%%%
\subsection{FEROS observations}
\label{feros}

%%%%%Table 4
\begin{table}
\caption{Log of the FEROS observations of XX~Sgr}
\begin{tabular}{c@{\hskip3mm}c@{\hskip3mm}c@{\hskip3mm}c@{\hskip3mm}r}
\hline
\noalign{\vskip 0.3mm}
Night       & Obs. ID. & Mid-exposure & Exposure & $v_{\rm rad}$\\
 & & JD 2\,400\,000+ & time (s) & (km/s)\\
\noalign{\vskip 0.3mm}
\hline
\noalign{\vskip 0.3mm}
15-16 Apr. &  F1487  & 55\,667.8105 & 180 & $-$1.98\\
16-17 Apr. &  F1645  & 55\,668.8165 & 180 &    6.04\\
17-18 Apr. &  F1723  & 55\,669.8192 & 180 &   15.00\\
18-19 Apr. &  F1805  & 55\,670.8533 & 180 &   22.48\\
\noalign{\vskip 0.3mm}
\hline
\noalign{\vskip 0.3mm}
\end{tabular}
\label{tab-ferosobs}
\end{table}

XX~Sgr was observed on four consecutive nights in April, 2011, 
using the FEROS spectrograph on the MPG/ESO 2.2\,m telescope 
in La Silla Observatory, Chile (see Table~\ref{tab-ferosobs} 
for details).
The Fiber-fed Extended Range Optical Spectrograph (FEROS)
(Kaufer et al.\,1999, 2000) has a total wavelength coverage
of 356-920\,nm with a resolving power of $R=48\,000$.
Two fibres, with entrance aperture of 2\farcs7, simultaneously
recorded star light and sky background. The detector is
a back-illuminated CCD with 2948$\times$4096 pixels of 15\,$\mu$m
size. Basic data reduction was performed using a pipeline package 
for reductions (DRS), in {\sc midas} environment. The pipeline performs 
the subtraction of bias and scattered light in the CCD, orders 
extraction, flatfielding and wavelength calibration with a ThAr 
calibration frame (the calibration measurements were performed 
at the beginning of each night, using the ThAr lamp).

After the continuum normalization of the spectra using 
{\sc iraf}\footnote{{\sc iraf} is distributed by the National Optical 
Astronomy Observatories, which are operated by the Association of 
Universities for Research in Astronomy, Inc., under cooperative 
agreement with the National Science Foundation.}
we determined the radial velocities with the task {\it fxcor\/}, 
applying the  cross-correlation method using a well-matching 
theoretical template spectrum from the extensive spectral library 
of \citet{M05}. The velocities were determined in the region 
500-600~nm where a number of metallic lines are present and lack of 
hydrogen lines. We made barycentric corrections to each radial 
velocity value with the task {\it rvcorrect\/}. The estimated 
uncertainty of the radial velocities is 0.05~km\,s$^{-1}$.

%%%%%%%%%%%%%
\subsection{Binarity of XX~Sgr}
\label{xxsgr-bin}

%%%Fig. 5
\begin{figure}
\includegraphics[height=55mm, angle=0]{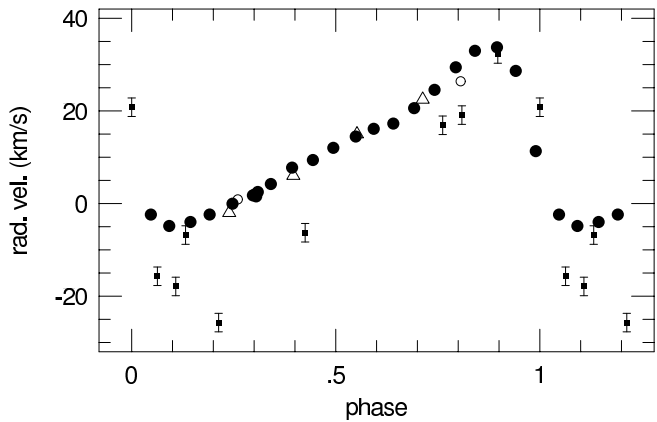}
\caption{Merged radial velocity phase curve of XX~Sgr.
There is a striking difference between the 
$\gamma$-velocities valid for the epoch of Joy's 
(\citeyear{J37}) data (denoted as small squares) and more 
recent data (other symbols, see the detailed list in the
text)}
\label{fig-xxsgr-vrad}
\end{figure}

All radial velocity data have been folded on the accurate
pulsation period taken from the ephemeris given in 
Eq.~\ref{xxsgr-ephemeris},
so the different data series have been correctly phased with
respect to each other. The merged radial velocity phase curve 
is plotted in Fig.~\ref{fig-xxsgr-vrad}.

The individual data series are denoted with different symbols:
small squares - \citet{J37} corresponding to mid-JD\,2\,426\,283; 
open circles - \citet{Betal88} with mid-JD\,2\,444\,254; 
filled circles - \citet{Setal11} and \citet{Betal10},
with mid-JD\,2\,454\,261 and mid-JD\,2\,454\,730, respectively;
triangles: our FEROS data listed in Table~\ref{tab-ferosobs}, 
with mid-JD\,2\,455\,669.

The earliest radial velocity data by \citet{J37} imply a
significantly different $\gamma$-velocity than all more recent
ones in spite of the uncertainty of his individual data as
large as 4 km\,s$^{-1}$. Because the zero point of Joy's system 
is reliable, as discussed by \citet{Sz96}, there is no systematic 
difference of instrumental or data treatment origin between Joy's 
and the more recent observational series. The only plausible 
explanation for the shift in the $\gamma$-velocity is the orbital 
motion in a binary system superimposed on the pulsational radial 
velocity changes.

Spectroscopic binarity of XX~Sgr can hardly be suspected from 
the radial velocity data obtained in the last 25 years, so the
orbital period is certainly much longer.

%%%%%%%%%%%%%%
\section{Conclusions}
\label{concl}

We pointed out that two bright Galactic Cepheids,
X~Puppis and XX~Sagittarii have a variable $\gamma$-velocity
which implies their membership in spectroscopic binary
systems. The available radial velocity data are
insufficient to determine the orbital period and
other elements of the orbit. It is obvious, however,
that the orbital period is as long as several decades
in both cases.
Such long orbital periods are not unprecedented among
classical Cepheids, cf. the cases of T~Mon and AW~Per 
\citep[see the on-line data base 
http://www.konkoly.hu/CEP/orbit.html
and its description in][]{Sz03a}.

Neither X~Pup, nor XX~Sgr shows any photometric
evidence of duplicity based on the criteria discussed
by \citet{Sz03b} and \citet{KSz09} indicating that the
companion cannot be a star much hotter than the 
Cepheid component in either case. Further spectroscopic 
observations are necessary to characterise these binary systems.

Regular monitoring of the radial velocities of a large
number of Cepheids will be instrumental in finding 
more long-period spectroscopic binaries among Cepheids. 
Quite recently \citet{Eetal12} reported on their 
on-going survey for pointing out binarity of Cepheids 
from the existing radial velocity data covering sufficiently 
long time interval.

When determining the physical properties (luminosity, 
temperature, radius, etc.) of individual Cepheids,
the effects of the companion on the observed parameters
(apparent brightness, colour indices, etc.) have to be
corrected for. This type of analysis, however, should be
preceded by revealing the binarity of the given Cepheid.

\section*{Acknowledgments}
%%%\begin{acknowledgements}
Financial support from the ESA PECS Project C98090, ESTEC
Contract No.4000106398/12/NL/KML, the Hungarian OTKA
Grants K76816 and K83790, the MB08C 81013 mobility grant,
and the ``Lend\"ulet-2009'' Young Researchers
Program of the Hungarian Academy of Sciences are gratefully acknowledged.
A. D. and Cs. K. acknowledges the support of the J\'anos Bolyai Research
Fellowship of the Hungarian Academy of Sciences.  A. D. was supported by 
the Hungarian E\"otv\"os fellowship. We are grateful to the staff at the 
2.2\,m telescope at La Silla for their kind support during the FEROS
observations. The {\it INTEGRAL\/} photometric data, 
pre-processed by ISDC, have been retrieved from the OMC Archive at CAB
\linebreak(INTA-CSIC). 
Critical remarks by Dr. M. Kun and the referee led to a considerable 
improvement in the presentation of the results.
%%%%\end{acknowledgements}

\bsp

\label{lastpage}

\end{document}